\input{aipcheck}

\documentclass[
    ,final            
  ]
  {aipproc}

\layoutstyle{6x9}

\begin{document}

\title{Relating Leptogenesis to Low Energy CP Violation}


\author{Mu-Chun Chen}{
  address={Theoretical Physics Department, Fermilab, Batavia, IL 60510, USA\\
  and\\
  Department of Physics \& Astronomy, University of California, Irvine, CA 92697, USA}
}

\author{K.T. Mahanthappa}{
  address={Department of Physics, University of Colorado, Boulder, CO 80309-0390, USA}
}

\begin{abstract}
 In the minimal left-right symmetric model with spontaneous CP violation, there are only two intrinsic CP violating phases to account for all CP violation in both the quark and lepton sectors. In addition, the left- and right-handed Majorana mass terms for the neutrinos are proportional to each other due to the parity in the model. This is thus a very constrained framework, making the existence of correlations among the CP violation in leptogenesis, neutrino oscillation and neutrinoless double beta decay possible. 
  \end{abstract}

\maketitle


\section{Introduction}

The evidence of non-zero neutrino masses opens up the possibility that the leptonic CP violation might be responsible, through leptogenesis, for the observed asymmetry between matter and anti-matter in the Universe. It is generally difficult, however, to make connection between leptogenesis and CP-violating processes at low energies due to the presence of extra phases and mixing angles in the right-handed neutrino sector. Recently attempts have been made to induce {\it spontaneous CP violation}  (SCPV)  from a single source.  Here we focus on the minimal left-right symmetric model. In this model SCPV could be due to two intrinsic CP violating phases associated with VEVs of two scalar fields which account for all CP-violating processes observed in Nature; these {\it exhaust} sources of CP-violation. As the left-handed (LH) and right-handed (RH) Majorana mass matrices are identical up to an overall mass scale, in this model there exist relations between low energy processes, such as neutrino oscillations, neutrinoless double beta decay and lepton flavor violating charged lepton decay, and leptogenesis which occurs at very high energy, as pointed out in Ref.~\cite{Chen:2004ww}.

\section{The Minimal Left-Right Symmetric Model}\label{model}

In the minimal left-right symmetric $SU(2)_{L} \times SU(2)_{R} \times 
U(1)_{B-L}$ model~\cite{Pati:1974yy}, the left- and right-handed 
matter fields transform as doublets of $SU(2)_{L}$ and $SU(2)_{R}$, 
respectively, and the Higgs sector that breaks the left-right symmetry 
to the SM gauge group contains a $SU(2)$ bi-doublet Higgs, $\Phi$, a 
$SU(2)_{L}$ triplet, $\Delta_{L}$, and a $SU(2)_{R}$ triplet, $\Delta_{R}$. 
The Yukawa interactions responsible for generating the lepton masses 
are, 
\begin{equation}
-\mathcal{L}_{\ell} =  
\overline{L}_{i,R} (P_{ij} \Phi + R_{ij} \tilde{\Phi}) L_{j,L} 
+ i f_{ij} (L_{i,L}^{T} \mathcal{C}\tau_{2} \Delta_{L} L_{j,L} 
+ L_{i,R}^{T} C\tau_{2} \Delta_{R} L_{j,R}) 
+ h.c. \; ,
\end{equation}
where  the matrices 
$P_{ij}$, $R_{ij}$ and $f_{ij}$ are real due to the assumption of SCPV.
The complete Lagrangian of the model is invariant under 
the unitary transformation, 
under which the matter fields transform as
$\psi_{L} \rightarrow U_{L} \psi_{L}$ and 
$\psi_{R} \rightarrow U_{R} \psi_{R}$
where $\psi_{L,R}$ are left-handed (right-handed) fermions, 
and the scalar fields transform according to
$\Phi \rightarrow U_{R} \Phi U_{L}^{\dagger}$,  
$\Delta_{L} \rightarrow U_{L}^{\ast} \Delta_{L} U_{L}^{\dagger}$, and 
$\Delta_{R} \rightarrow U_{R}^{\ast} \Delta_{R} U_{R}^{\dagger}$, 
with the unitary transformations $U_{L}$ and $U_{R}$ being 
\begin{equation}\label{unit}
U_{L}  =  
\left(
\begin{array}{cc}
e^{i\gamma_{L}} & 0\\
0 & e^{-i\gamma_{L}}
\end{array}
\right)
, \qquad 
U_{R} = 
\left(
\begin{array}{cc}
e^{i\gamma_{R}} & 0\\
0 & e^{-i\gamma_{R}}
\end{array}
\right) \; .
\end{equation}
Under these unitary transformations, the VEV's transform as
$\kappa  \rightarrow  \kappa e^{-i(\gamma_{L}-\gamma_{R})}$, 
$\kappa^{\prime} \rightarrow  \kappa^{\prime} e^{i(\gamma_{L}-\gamma_{R})}$, 
$v_{L}  \rightarrow  v_{L} e^{-2i\gamma_{L}}$ and 
$v_{R} \rightarrow  v_{R} e^{-2i\gamma_{R}}$. 
Thus by re-defining the phases of matter fields with the choice of  
$\gamma_{R}  =  \alpha_{R}/2$ and
$\gamma_{L} =   \alpha_{\kappa} + \alpha_{R}/2$ 
in the unitary matrices $U_{L}$ and $U_{R}$, 
we can rotate away two of the complex phases in the VEV's of 
the scalar fields and are left with only two genuine CP violating phases, 
$\alpha_{\kappa^\prime}$ and $\alpha_{L}$, 
\begin{equation}
<\Phi>  = \left(
\begin{array}{cc}
\kappa & 0\\
0 & \kappa^{\prime}e^{i\alpha_{\kappa^{\prime}}}
\end{array}
\right), \quad
<\Delta_{L}>  =  
\left(
\begin{array}{cc}
0 & 0 \\
v_{L}e^{i\alpha_{L}} & 0
\end{array}\right), \quad
<\Delta_{R}> = 
\left(
\begin{array}{cc}
0 & 0 \\
v_{R} & 0
\end{array}\right).
\end{equation}

The quark Yukawa interactions give rise 
to quark masses after the bi-doublet acquires VEV's
\begin{equation}
M_{u} = \kappa F_{ij} + \kappa^{\prime}  
e^{-i \alpha_{\kappa^\prime}} G_{ij}, 
\quad 
M_{d} = \kappa^{\prime} e^{i\alpha_{\kappa^\prime}}  F_{ij} 
+ \kappa G_{ij} \; .
\end{equation}
Thus the relative phase in the two VEV's in the SU(2) 
bi-doublet, $\alpha_{\kappa^\prime}$, gives rise 
to the CP violating phase in the CKM matrix. 
To obtain realistic quark masses and CKM matrix elements, 
it has been shown that the VEV's of the bi-doublet 
have to satisfy $\kappa/\kappa^\prime \simeq m_{t}/m_{b} \gg 1$.
When the triplets and the bi-doublet acquire VEV's, we obtain the following 
mass terms for the leptons
\begin{eqnarray}
M_{e} = \kappa^{\prime} e^{i\alpha_{\kappa^\prime}} P_{ij} + \kappa R_{ij}, 
& \quad
M_{\nu}^{Dirac} = \kappa P_{ij} 
+ \kappa^{\prime} e^{-i\alpha_{\kappa^\prime}} R_{ij} \\
M_{\nu}^{RR} = v_{R} f_{ij}, & \quad
M_{\nu}^{LL} = v_{L} e^{i\alpha_{L}} f_{ij} \; .
\end{eqnarray}
The effective neutrino mass matrix, $M_{\nu}^{\mbox{\tiny eff}}$, which 
arises from the Type-II seesaw mechanism,  
is thus given by $M_{\nu}^{\mbox{\tiny eff}} = M_{\nu}^{LL} - M_{\nu}^{I}$. 
Assuming the charged lepton mass matrix is diagonal, 
the Yukawa couplings $R_{ij}$ can be determined  
by the charged lepton masses. In the limit $\kappa \gg \kappa^\prime$, the 
conventional type-I see-saw term  
is dominated by the term proportional to $\kappa$,
\begin{equation}
M_{\nu}^{I} = (M_{\nu}^{\mbox{\tiny Dirac}})^{T} 
(M_{\nu}^{RR})^{-1} (M_{\nu}^{\mbox{\tiny Dirac}})
\simeq 
\frac{\kappa^{2}}{v_{R}}P^{T} f^{-1} P 
= \frac{v_{L}}{\beta}P^{T} f^{-1} P
\; .
\end{equation}

\section{Results}

In the left-right symmetric model with the particle content we 
have, leptogenesis receives contributions both from 
the decay of the lightest RH neutrino, $N_{1}$, as well as 
from the decay of the $SU(2)_{L}$ triplet Higgs, 
$\Delta_{L}$. 
We consider the $SU(2)_{L}$ triplet Higgs 
being heavier than the lightest RH neutrino, $M_{\Delta_{L}} > M_{R_{1}}$. 
For this case, the decay of the lightest RH neutrino dominates.
In the SM, the canonical contribution to the lepton number 
asymmetry from one-loop diagrams 
mediated by the Higgs doublet and the charged leptons 
is given by~\cite{Joshipura:2001ya} ,    
\begin{equation}
\epsilon^{N_{1}}  =  
\frac{3}{16\pi }  \biggl( \frac{M_{R_{1}}  }{v^{2}} \biggr)  
\cdot \frac{
\mbox{Im} \biggl(  \mathcal{M}_{D}  \left( M_{\nu}^{I} \right)^{\ast} 
\mathcal{M}_{D}^{T} 
\biggr)_{11} }{ ( \mathcal{M}_{D} \mathcal{M}_{D}^{\dagger} )_{11} }  \; . 
\end{equation}
Now, there is one additional one-loop diagram mediated 
by the $SU(2)_{L}$ triplet Higgs.  It contributes to the 
decay amplitude of the right-handed neutrino into a doublet 
Higgs and a charged lepton, which 
gives an additional contribution to the lepton number 
asymmetry~\cite{Joshipura:2001ya},   
 \begin{equation}
 \epsilon^{\Delta_{L}}  =  
\frac{3}{16\pi }  \biggl( \frac{M_{R_{1}}  }{v^{2}} \biggr)  
\cdot \frac{
 \mbox{Im} \biggl(  \mathcal{M}_{D}  \left( M_{\nu}^{II} \right)^{\ast} 
\mathcal{M}_{D}^{T} 
 \biggr)_{11} }{ ( \mathcal{M}_{D} \mathcal{M}_{D}^{\dagger} )_{11} }  \; ,
 \end{equation} 
where $\mathcal{M}_{D}$ is the neutrino Dirac mass term in the basis where 
the RH neutrino Majorana mass term is real and diagonal, 
$\mathcal{M}_{D} = O_{R} M_{D}$ and  
$f^{\mbox{\tiny diag}} = O_{R} f O_{R}^{T}$. 
Because there is no phase present in either 
$M_{D} = P \kappa$ or $M_{\nu}^{I}$ or $O_{R}$, 
the quantity $ \mathcal{M}_{D} \left( M_{\nu}^{I} \right)^{\ast} 
\mathcal{M}_{D}^{T}$ is real, leading to a vanishing $\epsilon^{N_{1}}$. 
We have checked explicitly that this statement is 
true for {\it any} chosen unitary transformations $U_{L}$ and $U_{R}$ 
defined in Eq.~(\ref{unit}).  
On the other hand, the contribution, 
$\epsilon^{\Delta_{L}}$, due to the diagram mediated by 
the $SU(2)_{R}$ triplet is proportional to $\sin\alpha_{L}$. 
So, as long as the phase $\alpha_{L}$ is non-zero, 
the predicted value for $\epsilon^{\Delta_{L}}$ is finite.   
A non-vanishing value for 
$\epsilon^{N_{1}}$ is generated at the sub-leading order when 
terms of order $\mathcal{O}(\kappa^{\prime}/\kappa)$ in $M_{D}$ are included. 
At the leading order, leptogenesis is generated solely from the decay 
mediated by the $SU(2)_{L}$ triplet Higgs.

Assume the neutrino Yukawa coupling $P_{ij}$ to be proportional to the up quark mass matrix and the matrix $f_{ij}$ having hierarchical elements,
\begin{equation}
P_{ij} = q \left(
\begin{array}{ccc}
\frac{m_{u}}{m_{t}} & 0 & 0 \\
0 & \frac{m_{c}}{m_{t}} & 0 \\
0 & 0 & 1
\end{array}
\right) \; , \quad f_{ij} = \left(\begin{array}{ccc}
t^{2} & t & -t\\
t & 1 & 1\\
-t & 1 & 1
\end{array}\right) \; , 
\end{equation}
with $q$ being the proportionality constant and $t$ a small and positive number. 
CP violation in neutrino oscillation is governed by the leptonic Jarlskog invariant, 
$J_{\mbox{\tiny CP}}$, 
which can be expressed model-independently 
in terms of the effective neutrino mass matrices as 
$J_{\mbox{\tiny CP}} = 
-\mbox{Im}(H_{12}H_{23}H_{31})/(\Delta m_{21}^{2} 
\Delta m_{31}^{2} \Delta m_{32}^{2})$ and  
$H \equiv M_{\nu}^{\mbox{\tiny eff}} M_{\nu}^{\mbox{\tiny eff}\dagger}$, 
where $\Delta m_{ij}^{2} \equiv m_{i}^{2} - m_{j}^{2}$ with $m_{i}$ being the 
mass eigenvalues of the effective neutrino mass matrix, 
$M_{\nu}^{\mbox{\tiny eff}}$. 

To the leading order in $\epsilon$ 
the leptonic Jarlskog invariant is given by,
\begin{equation}
J_{\mbox{\tiny CP}} 
\simeq -\frac{2st^{2}\left(1-t^{2}\right)v_{L}^{6}}{
\Delta m_{21}^{2} \Delta m_{31}^{2} \Delta m_{32}^{2}}
\frac{m_{u}}{m_{t}} \sin\alpha_{L} \; ,
\end{equation}
and the total amount of lepton number asymmetry, 
$\epsilon_{\mbox{\tiny total}} = \epsilon^{\Delta_{L}}$,  
is proprtional to $\Delta \epsilon^{\prime}$, defined as
\begin{equation}
\Delta \epsilon^{\prime} 
= \frac{3}{16\pi} \frac{f_{1}^{0}}{v_{L}}  \cdot  
 \frac{
 \mbox{Im} \biggl(  \mathcal{M}_{D}  \left(v_{L} f \, e^{i\alpha_{L}} 
\right)^{\ast} 
\mathcal{M}_{D}^{T} 
 \biggr)_{11} }{ ( \mathcal{M}_{D} \mathcal{M}_{D}^{\dagger} )_{11} } 
= \frac{\epsilon^{\Delta_{L}}}{\beta} \propto \sin\alpha_{L} \; ,
\end{equation}
where $\beta$ is a function of order one coupling constants in the scalar potential. 
The leptonic Jarlskog $J_{\mbox{\tiny}}$ and the amount of leptogenesis $\epsilon^{\Delta_{L}}$ are both 
proportional to $\sin\alpha_{L}$. Thus the observation of low energy CP violation then implies a non-vanishing lepton number asymmetry. 
%
We note that an amount of lepton number asymmetry 
$\epsilon_{\mbox{\tiny total}} \sim 10^{-8}$ corresponds to 
a leptonic Jarlskog invariant $|J_{\mbox{\tiny CP}}| \sim 10^{-5}$.  

With an additional broken $U(1)$ symmetry, the seesaw scale can be lowered to be closer to the EW scale. In that case, the phase $\alpha_{\kappa^{\prime}}$ responsible for  CPV in the quark sector can also be relevant to the leptonic CP violating processes. Connections such as the one  between electron EDM and leptogenesis can thus be established~\cite{Chen:2006bv}.


\begin{figure}
 \includegraphics[height=.3\textheight,angle=270]{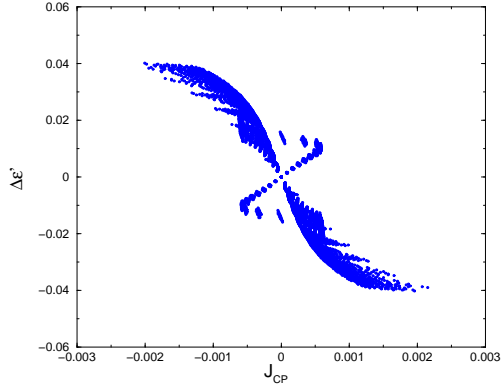}
 \caption{Correlation between lepton number asymmetry and the leptonic Jarlskog invariant.}
\end{figure}




\bibliographystyle{aipproc}   



\begin{thebibliography}{9}

\bibitem{Chen:2004ww}
  M.-C.~Chen and K.~T.~Mahanthappa,
  \emph{Phys.  Rev.} \textbf{D71}, 035001 (2005).


\bibitem{Pati:1974yy}
  J.~C.~Pati and A.~Salam,
 \emph{Phys. Rev.} \textbf{D10}, 275 (1974); 
  R.~N.~Mohapatra and J.~C.~Pati,
 \emph{Phys. Rev.} \textbf{D11}, 566 (1975). 


\bibitem{Joshipura:2001ya}
  A.~S.~Joshipura, E.~A.~Paschos and W.~Rodejohann,
  \emph{Nucl. Phys.} \textbf{ B611}, 227 (2001).
  S.~Antusch and S.~F.~King,
  Phys.\ Lett.\ B {\bf 597}, 199 (2004).
  

\bibitem{Chen:2006bv}
  M.-C.~Chen and K.~T.~Mahanthappa,
  hep-ph/0609288.


\end{thebibliography}





\end{document}